\documentclass[letterpaper, 10 pt, conference]{ieeeconf}  

\IEEEoverridecommandlockouts                              
\usepackage{graphics} 
\usepackage{epsfig} 
\usepackage{mathptmx} 
\usepackage{times} 
\usepackage{amsmath} 
\usepackage{amssymb}  

\title{\LARGE \bf
Handling-Oriented Stiffness Control of a Multichamber Suspension
}

\author{Gabriele Marini, Giulio Panzani, Matteo Corno, Samuele Sermisoni, Sergio M. Savaresi
\thanks{Authors are with the Department of Electronics, Information and Bioengineering, Politecnico di Milano, Piazza L. Da Vinci 32, Milano, 20133, Italy
        {\tt\small gabriele.marini@polimi.it}}%
}

\begin{document}

\maketitle
\thispagestyle{empty}
\pagestyle{empty}

\begin{abstract}
This paper deals with the development of a handling-oriented stiffness control strategy using multichamber suspensions. Indeed, being this technology capable of stiffness variability, it is particularly indicated for improving the vehicle handling performance, here intended as the reduction of roll and pitch angles during maneuvers. The proposed strategy exploits the multichamber's inner features in order to enhance the performance: simulation results show improvements up to 12\% compared to the best passive stiffness configuration, still preventing deterioration of the driving comfort. 
\end{abstract}

\section{Introduction}
In road vehicles, suspension systems have an all-round effect on the feeling perceived during the ride. They indeed are the main responsible for ensuring comfort, safety and handling performance, here intended as the reduction of the vehicle roll and pitch rotations during maneuvers. Passive suspensions are affected by a trade-off wherein stiffer suspensions yield a more stable response to the driver inputs (better handling) but are not effective in filtering road disturbances (worse sense of comfort); and vice versa for softer suspensions \cite{SAVARESI2010}\cite{NING2010}.

The compromise between handling and ride comfort is partially overcome relying on architectures that allow us to modulate their stiffness and damping properties. Several technologies are available: variable-damping shock-absorbers \cite{CORNO2019667}\cite{LU2002}, which however reduce dynamical oscillations and not steady-state angles during maneuvers; slowly adaptive pneumatic suspensions with load-levelling capability \cite{KIM2011}, which only compensate for static load variations; active suspensions, which are the most performing ones, however resulting in high-power consumption and fault tolerance issues \cite{JIN2016}\cite{GOHRLE2014}.

Nowadays, one promising architecture with dynamical stiffness modulation capability for reduced steady-state rolling and pitching is the multichamber air suspension \cite{DATTILO2020}. In these suspensions, a set of auxiliary air reservoirs can be attached to (or detached from) the main pneumatic chamber by means of controllable valves, thus resulting in a change of the total spring stiffness. In recent years, the multichamber technology has been exploited for handling purposes on some high-end production cars, thanks to the reduced cost and energy demand which valve modulation implies. These systems usually permit manual setting of the desired driving style, whereas only a few manufacturers report automatic soft-to-hard regulation during handling maneuvers \cite{GANTIKOW2017}\cite{MERCEDES}. However, to the best of the authors’ knowledge, no documentation on this specific topic has ever been provided in the literature.

This paper deals with the development of a handing-oriented control technique specifically suited for a multichamber suspension. The proposed approach exploits this technology’s features to enhance handling performance with respect to the basic soft-to-hard switching logic. Moreover, it makes use of an automatic maneuver recognition policy based on the longitudinal and lateral acceleration. The strategy is also suitable for real time application, thanks to its efficient \textit{if-then-else} formulation. Simulations, conducted on a full vehicle simulator, show an overall improvement of the vehicle angular rotations of 12\%, while avoiding a deterioration of the vehicle vertical acceleration that may arise from valve switching.

This paper is organized as follows. Section II describes the simulation model; Section III analyzes the passive handling performance; Section IV and V describe the maneuver detection and the control strategies; Section VI  and VII contain simulation results and conclusions.

\section{Simulation model}
This section describes the simulation model used for handling-oriented control.

\subsection{Suspension model}
\begin{figure}[b]
	\begin{center}	
\includegraphics[width=0.35\textwidth]{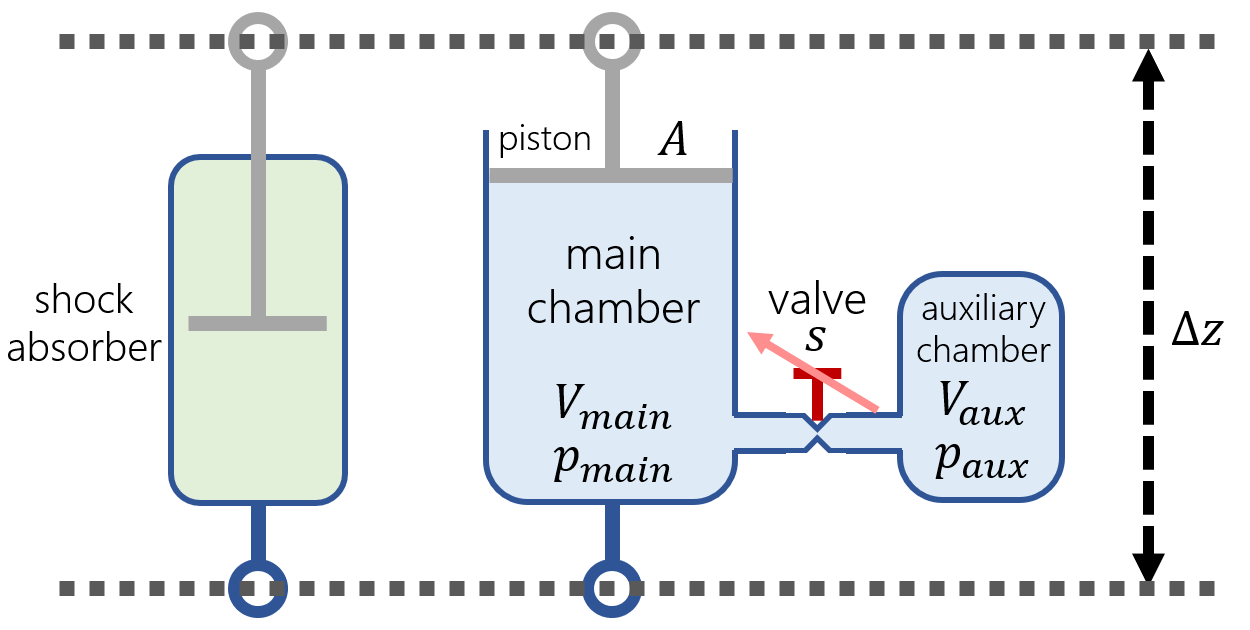}
	\caption{Schematic representation of a multichamber suspension with one auxiliary chamber.}
	\label{fig:MC_susp}
	\end{center}
\end{figure}

A suspension is composed of the parallel of a spring and shock-absorber. In the case of a multichamber suspension, the elastic element is made of a main pneumatic chamber attached to one or more auxiliary air chambers. The main chamber has variable volume, following the piston movement, whereas the auxiliary volumes are fixed. Without loss of generality, the case of one auxiliary reservoir is considered (see Fig. \ref{fig:MC_susp}).

The auxiliary chamber is connected to the main one via a controllable valve, whose on-off state regulates the amount of volume $V$ subject to compression during the ride. In particular, the stiffness coefficient $k$ of a classic pneumatic spring depends on the total air volume, as in:
\begin{equation}
k=\frac{\gamma \bar{p} A^2}{V},
\label{k_MC}
\end{equation}
where $\gamma$ is the air polytropic coefficient, $\bar{p}$ the equilibrium pressure and $A$ the piston area. Eq. \eqref{k_MC} tells that changing volume via valve switching leads to the regulation of the spring stiffness; the higher the volume, the softer the spring, and viceversa.

\begin{figure}[t]
	\begin{center}	
\includegraphics[width=0.48\textwidth]{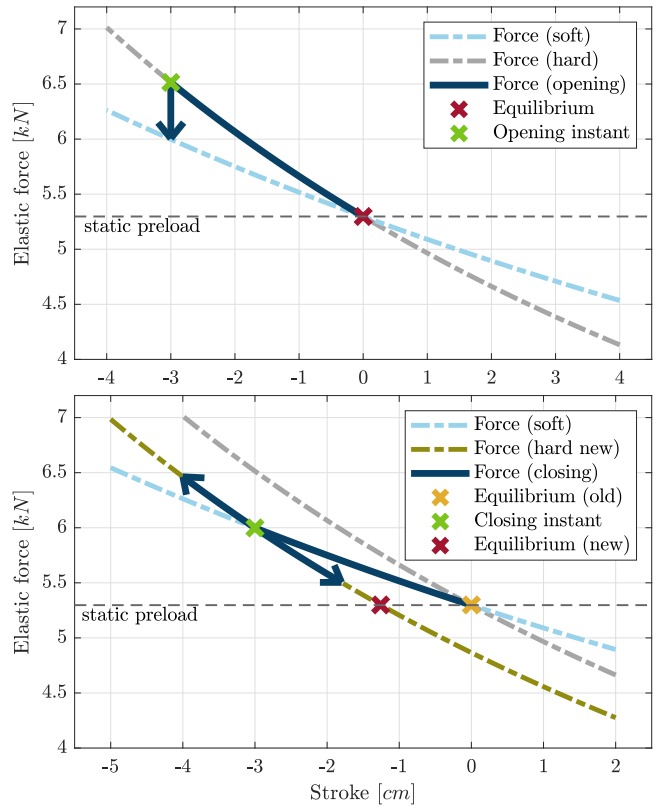}
	\caption{Stroke-force maps for different valve configurations (dashed lines). The arrows show the evolution of the suspension force after opening (upper plot) and closing (lower plot) the valve.}
	\label{fig:switching_maps_vert}
	\end{center}
\end{figure}

This paper makes use of the thermodynamical model described in \cite{DATTILO2020}. It relates the stroke movement to the internal pressure of the chambers, and hence to the elastic force exerted by the spring. The model inputs are the discrete valve position $s\in\{0-closed;1-open\}$ (controllable) and the stroke elongation $\Delta z$ (exogenous). Fig. \ref{fig:switching_maps_vert} shows the behaviour of the multichamber spring by means of its elastic maps, whose main features are highlighted in the following.

\begin{itemize}
\item \textit{Stiffness variability.} Fig. \ref{fig:switching_maps_vert} (top) reports the soft and hard stiffness maps, obtained by keeping the valve open and closed respectively. As opposed to a linear spring, the elastic maps are \textit{progressive} in compression and \textit{regressive} in extension, meaning that the force variation at a given stroke variation is higher in compression than in extension.
\item \textit{Valve opening.} When the valve opens, the elastic force jumps from the hard map to the soft one (Fig. \ref{fig:switching_maps_vert} (top)). This jump is called \textit{kick force}, and is given by the pressure drop of the two unequally-pressurized chambers.
\item \textit{Valve closing.} When the valve closes, the elastic force switches to the hard map configuration without any kick force, resulting in a \textit{shifted} replica of the original hard map (Fig. \ref{fig:switching_maps_vert} (bottom)). As an important consequence, the stroke equilibrium position changes, that means the suspension stroke is shifted to a new position in order to compensate for the same static vertical load. In order to go back to the original equilibrium position, the valve must be opened again.
\end{itemize}

The shock-absorber, that completes the suspension item, is modelled as a static element with the force $F_c$ being proportional to the stroke speed $\Delta \dot{z}$ through the damping coefficient $c$: 
\begin{equation}
F_{c}=-c\Delta \dot{z}.
\end{equation}
For the complete model equations, the reader is referred to \cite{DATTILO2020}.

\subsection{Vehicle model}
The vehicle model used in this work is a multi-body sedan-type car model, defined in the \textit{VI-Grade} simulation environment. The standard (passive) suspension forces are substituted and defined externally, according to the previously presented multichamber architecture, so that stiffness can be modulated at the four corners.

The main parameters involved in simulation are reported in Table \ref{table:param} (end of this paper). The suspension parameters are those of commercially available multichamber suspensions. For realistic simulation purposes, a minimum switching interval of 100ms is introduced, in order to account for the physical switching time of the valve.

\section{Handling-oriented analysis}

\begin{figure}[t]
	\begin{center}	
\includegraphics[width=0.48\textwidth]{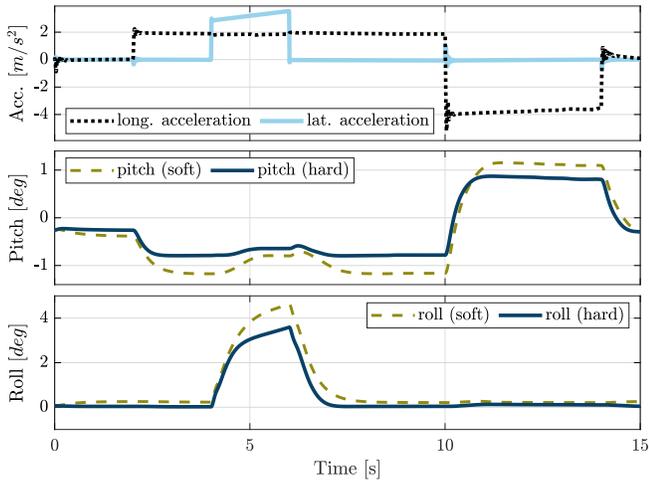}
	\caption{Example of roll and pitch angles during mixed handling maneuver.}
	\label{fig:LTransfer_example}
	\end{center}
\end{figure}

This work focuses on the vehicle handling performance, hereby defined as the steady-state pitch and roll angles achieved during maneuvers that involve throttle, brake and steering actions. Such maneuvers are hence referred to as \textit{handling maneuvers}. During these, the vehicle attitude (except for the yaw rotation) strictly depends on the suspension strokes. Therefore, ensuring limited stroke elongations at the four corners leads to a decrease of the pitch and roll rotations. This can be simply achieved by using the harder suspension configuration, as shown in Fig. \ref{fig:LTransfer_example}, where the soft and the hard configurations are compared, during a combination of longitudinal and lateral maneuvers.

However, when addressing the driving comfort, \textit{i.e.} the vehicle vertical movements due to road unevenness, the hard suspension configuration is well-known to be worse than the soft one \cite{SAVARESI2010}. In this paper, we assume the suspension to be normally set soft; the control strategy described in the following properly manages the transition to the hard configuration whenever a maneuver is detected.

\section{Maneuver recognition}
This section proposes a maneuver recognition technique for handling purposes.

\subsection{Corner load transfer}
\begin{figure}[thpb]
	\begin{center}	
\includegraphics[width=0.41\textwidth]{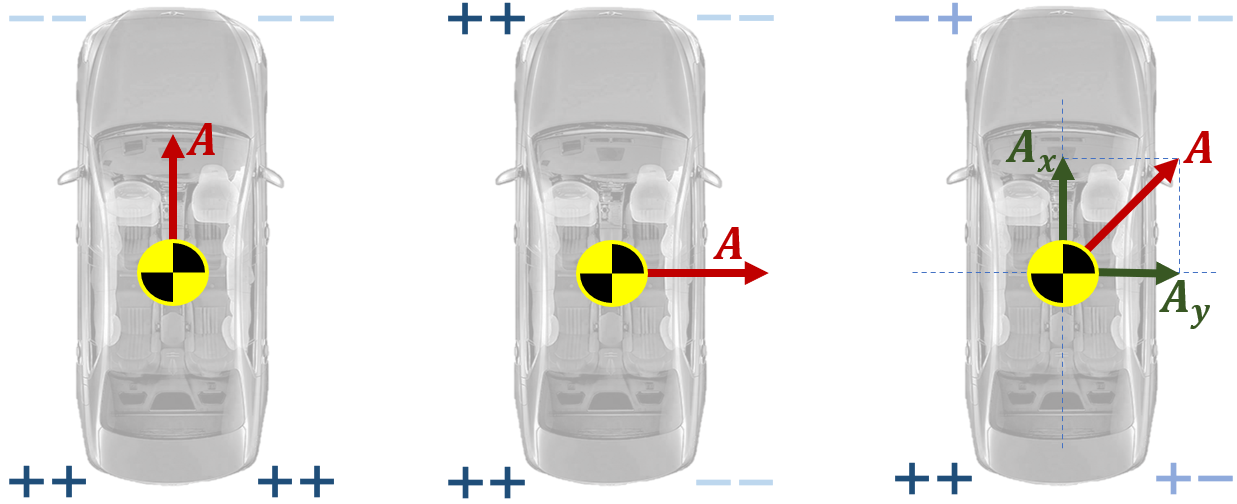}
	\caption{Corner load variations during single and mixed maneuvers; `$+$' means load is added, `$-$' means load is removed.}
	\label{fig:mix_transfer}
	\end{center}
\end{figure}

A signal for maneuver detection must be able to promptly identify the beginning of handling maneuvers. In general, they can be recognized starting from the longitudinal and lateral accelerations of the vehicle. Indeed, as seen in Fig. \ref{fig:mix_transfer} (left and middle), these accelerations redistribute the loads among corners, thus affecting the status of the suspensions. In particular, during traction (braking), load is transferred front to rear (rear to front); during left (right) steering, load is transferred right to left (left to right). When load is removed from a corner, the suspension elongates; when added, the suspension compresses. In these simple cases, longitudinal and lateral accelerations can hence be independently used as a way to detect the stroke status.

A more complex situation is represented by mixed maneuvers, where it is not possible to predict the status of two symmetrically placed corners relying on the measure of acceleration only. For example, as Fig. \ref{fig:mix_transfer} (right) shows, during a mixed steering/traction maneuver, the front-left end rear-right corners are both affected by load transfers having opposite \textit{directions}. In this case, the corresponding suspension behaviour will hence depend on the resulting load variation that each corner experiences.

\subsection{Simplified load transfer model}
The load transfer at each corner is obtained by considering the longitudinal and lateral inertial forces at the COG and neglecting the static chassis weight as well as the force dynamical contributions given by vehicle aerodynamics and road-induced stroke oscillations. 

\begin{figure}[thpb]
	\begin{center}	
\includegraphics[width=0.48\textwidth]{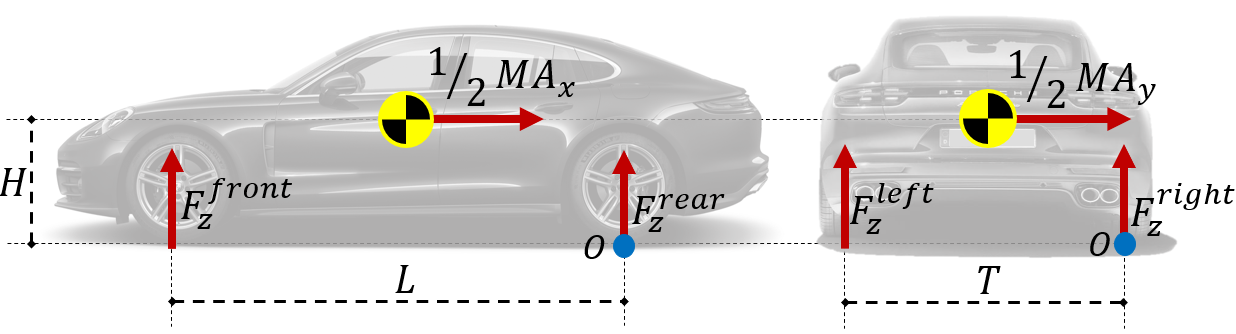}
	\caption{Single track (left) and single axle (right) models.}
	\label{fig:LTransfer}
	\end{center}
\end{figure}

A simple longitudinal load transfer model can be obtained assuming the vehicle mass entirely placed at the COG point, and constant COG height. The relation between the load variation $F_z$ at front and rear corners and the longitudinal acceleration $A_x$ is obtained via equilibrium of forces and torques (around point $O$) using a single track model with half the vehicle mass, as in Fig. \ref{fig:LTransfer} (left). For simplicity, the contribution of longitudinal and lateral tyre forces is not reported, since they don't affect the pitching torque around $O$. The same is done for the lateral transfer if one considers a single axle of the vehicle (Fig. \ref{fig:LTransfer} (right)), with its COG subject to lateral acceleration $A_y$. Also in this case, longitudinal and lateral tyre forces do not contribute to the rolling torque around $O$. 

By applying the superposition principle, the load variations obtained with the single maneuvers are summed, thus resulting in the following linear matrix equality:
\begin{equation}
\label{load_transfer}
\begin{bmatrix}
F_z^{FL} \\ F_z^{FR} \\ F_z^{RL} \\ F_z^{RR}
\end{bmatrix}
=\begin{bmatrix}
-\frac{HM}{2L} & -\frac{HM}{2T} \\ -\frac{HM}{2L} & +\frac{HM}{2T} \\ +\frac{HM}{2L} & -\frac{HM}{2T} \\ +\frac{HM}{2L} & +\frac{HM}{2T}
\end{bmatrix}
\begin{bmatrix}
A_x \\ A_y
\end{bmatrix},
\end{equation}
where the superscripts in the $F_z$'s indicate Front/Rear and Left/Right corners. The model in \eqref{load_transfer} exclusively relies on the values of the total mass $M$ and of the vehicle structural parameters, namely wheelbase $L$, track $T$ and COG height $H$. 

\begin{figure}[thpb]
	\begin{center}	
\includegraphics[width=0.48\textwidth]{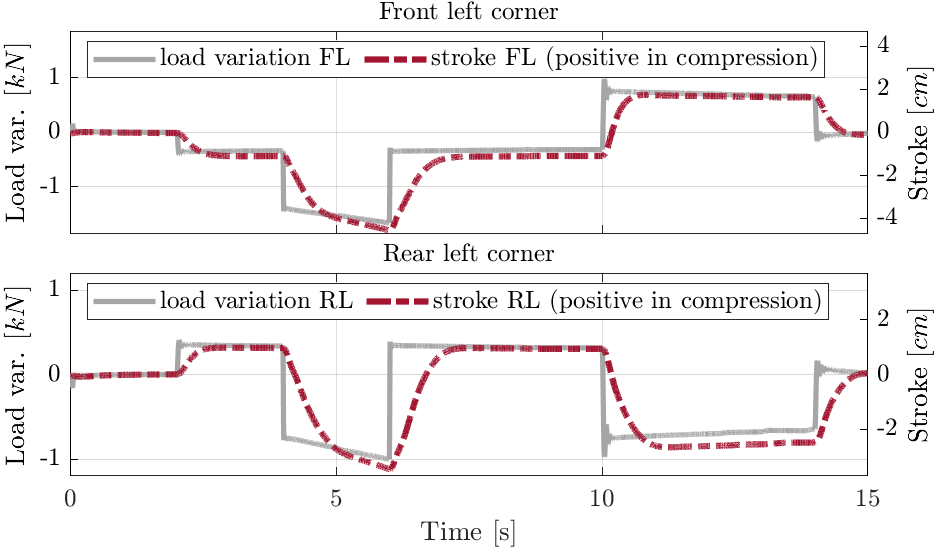}
	\caption{Load transfers and stroke movements during mixed maneuvers.}
	\label{fig:LTransfer_strokes}
	\end{center}
\end{figure}

An estimation example is reported in Fig. \ref{fig:LTransfer_strokes}. Following the same mixed maneuver considered in the previous section, each corner has its own load distribution (left \textit{y}-axis) depending on the intensity and direction of the acceleration vector in time. As a consequence, the suspension stroke (right \textit{y}-axis) reacts according to the load variation. Also, the higher the acceleration intensity, the longer the stroke. Due to the filtering action exerted by the system, the stroke movement is slower than the load transfer dynamics (and hence the acceleration dynamics); thanks to that, longitudinal and lateral accelerations alone represent suitable indexes to \textit{predict} and \textit{anticipate} the suspension behaviour. Relation \eqref{load_transfer} therefore represents a prompt and accurate way for the maneuver recognition, and can be used as a switching threshold for the handling-oriented control logic.

\section{Handling-oriented control}

This section describes the main concept of handling-oriented control, and introduces an innovative strategy which exploits the multichamber technology's inner features.

\subsection{Basic hardening logic}
\begin{figure}[thpb]
	\begin{center}	
\includegraphics[width=0.47\textwidth]{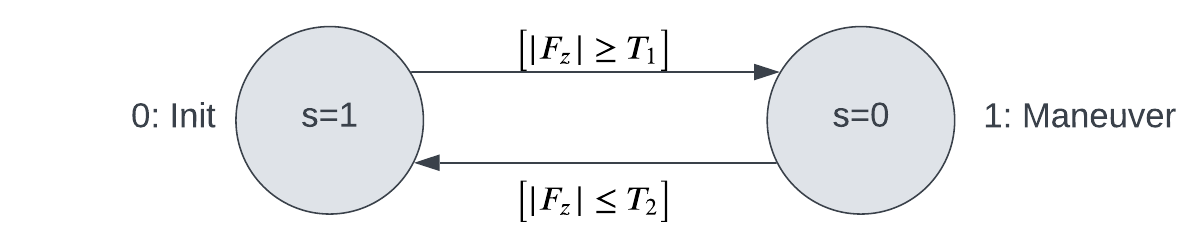}
	\caption{Core logic at corner $ij\in\{FL,FR,RL,RR\}$. Conventionally, $F_z:=F_z^{ij}$ and $s:=s_{ij}$.}
	\label{fig:V1}
	\end{center}
\end{figure}

The handling-oriented analysis highlights the necessity of switching to hard mode whenever a maneuver is detected, and back to soft mode after it has ended. Using a multichamber spring, the previous logic is easily translated into the one depicted in Fig. \ref{fig:V1}. Whenever the absolute value of the load transfer force exceeds a positive \textit{closing} threshold $T_1$, the valve closes thus stiffening the spring and reducing the total suspension stroke. This logic is applied independently at the four vehicle corners (\textit{decentralized} approach); in this way, the overall pitch and roll rotations benefit as a consequence of the reduced strokes. The opening of the valve is ruled by an \textit{opening} threshold $T_2$ on the load force, which states the end of the maneuver. 

The closing threshold can be tuned based on a \textit{critical acceleration} level, using relation \eqref{load_transfer}. This method is suitable in case one may not want to harden the stiffness configuration for low acceleration values (\textit{light} maneuvers). Moreover, closing and opening thresholds are generally different, to avoid chattering of the control signal in presence of disturbances and measurement noise. 

The proposed rationale applies equally to the case of a generic variable-stiffness spring, not necessarily achieved with the multichamber architecture.

\subsection{A multichamber-oriented switching logic}
\begin{figure}[b]
	\begin{center}	
\includegraphics[width=0.50\textwidth]{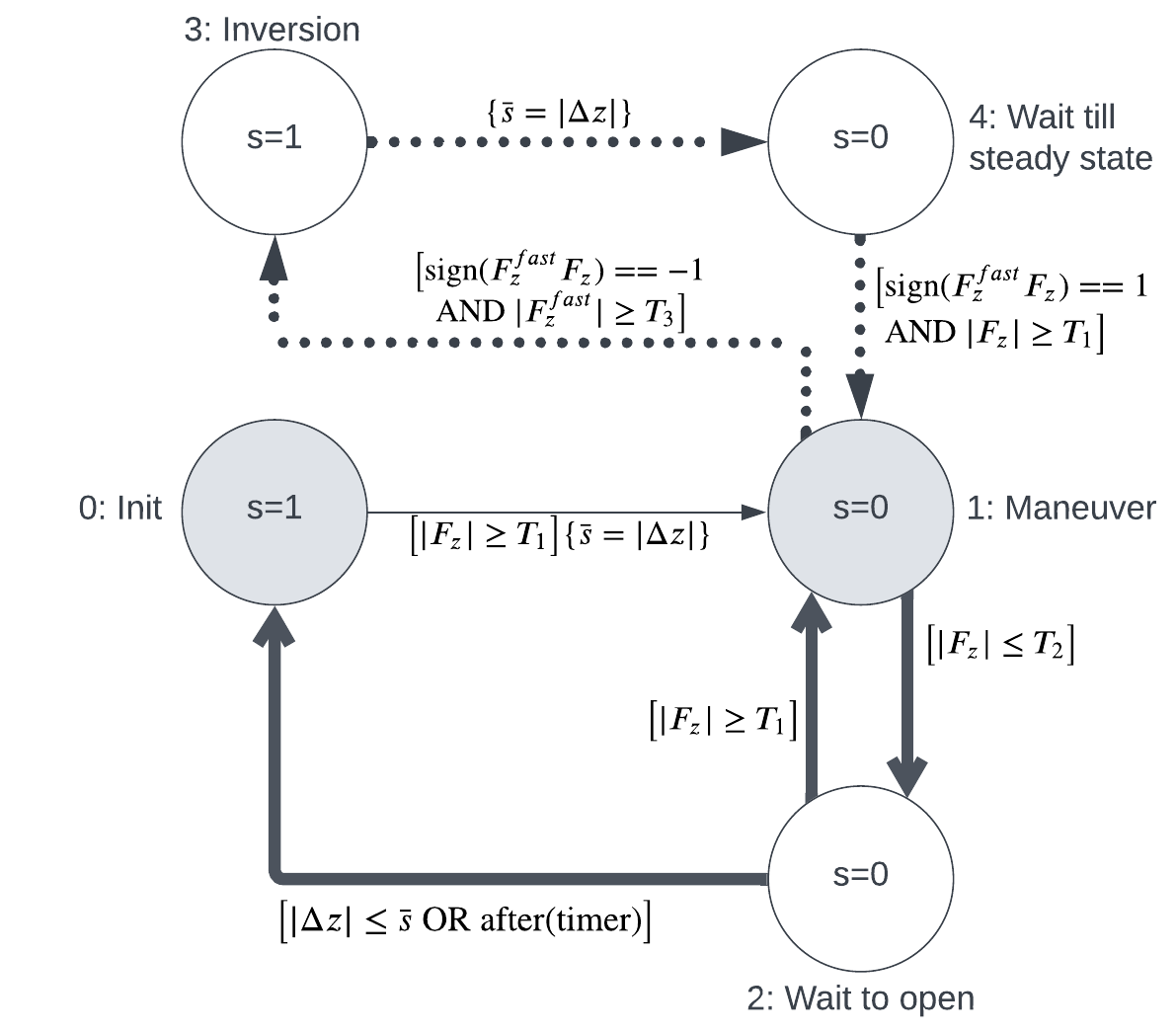}
	\caption{Handling controller. States n. 0 and 1 are those present in the core logic. Same nomenclature holds.}
	\label{fig:V3}
	\end{center}
\end{figure}

A multichamber suspension has intrinsic features both at valve opening and closing that must be specifically addressed and that can be exploited in order to enhance the handling performance achievable by the above described hardening logic. For this reason, this work proposes an extended control strategy, summarized in Fig. \ref{fig:V3}. It is composed of additional states which are grouped into two \textit{cycles} (highlighted with different line styles); each cycle implements a switching policy, tailored for the specific goals described in the following.
\begin{figure*}[t]
	\begin{center}	
\includegraphics[width=0.95\textwidth]{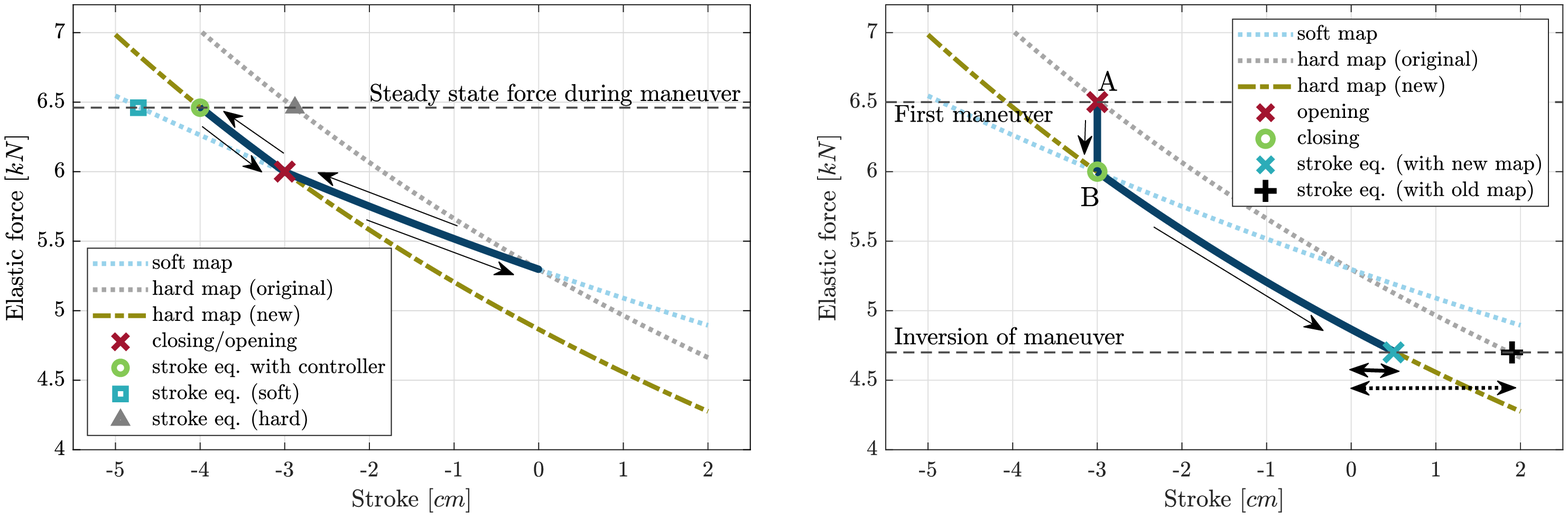}
	\caption{Visualization of kick force avoidance (left) and inversion of maneuver (right) using elastic maps.}
	\label{fig:visualizations}
	\end{center}
\end{figure*}
\\

\noindent(i) \textit{Kick force avoidance.} The cycle in bold line manages the opening policy in order to mitigate the kick-force effect, which is detrimental to the driving comfort. To do so, once the opening threshold is crossed (condition $|F_z|\leq T_2$), the actual valve opening is postponed until the moment when the stroke equals the stroke level of the previous closing instant (condition $|\Delta z|\leq \bar{s}$). In this way, opening occurs only when the pressures inside the chambers are the same, thus avoiding the generation of kick-force. This principle can effectively be visualized via elastic maps (see Fig. \ref{fig:visualizations} (left)). 

Alternatively, if the same stroke level is not reached again, a timing condition is inserted as a \textit{backup} opening strategy. Indeed, this last condition can become true due to the road stochasticity, which may prevent the stroke from crossing its previous closing level at the end of a maneuver.
\\

\noindent (ii) \textit{Maneuver inversion management.} The cycle in dotted line manages the cases where an \textit{inversion} of maneuver occurs. An inversion is defined as a maneuver where the load transfer of a corner inverts its sign (and keeps relatively large in modulus, by imposing a threshold $T_3$). As an example, in Fig. \ref{fig:LTransfer_strokes} the FL corner undergoes one inversion, whereas the corner RL experiences three. An inversion is detected through the condition $\text{sign}(F_z^{fast}F_z)==-1$, where $F_z^{fast}$ and $F_z$ are obtained filtering the load transfer force with a high-bandwidth (\textit{fast} signal) and low-bandwidth (\textit{slow} signal) low-pass filter respectively. 

This logic states that, once an inversion is detected, the valve gets promptly opened and closed again, in order to change equilibrium position and reduce the stroke at steady state during the inverted maneuver. An example that visualizes this physical principle is given by Fig. \ref{fig:visualizations} (right). Here, it is schematically shown that the steady-state stroke using maneuver inversion management (full double arrow) is lower than the one obtained using the original hard map (dotted double arrow).

\section{Simulation results}
\begin{figure}[b]
	\begin{center}	
\includegraphics[width=0.48\textwidth]{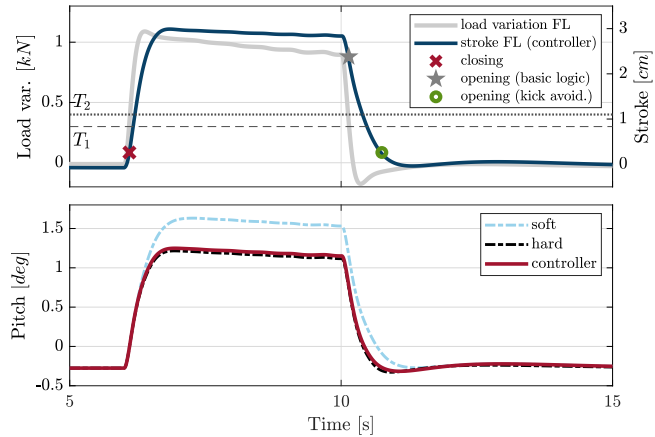}
	\caption{Kick force avoidance example.}
	\label{fig:kick_avoid_example}
	\end{center}
\end{figure}
\begin{figure}[b]
	\begin{center}	
\includegraphics[width=0.48\textwidth]{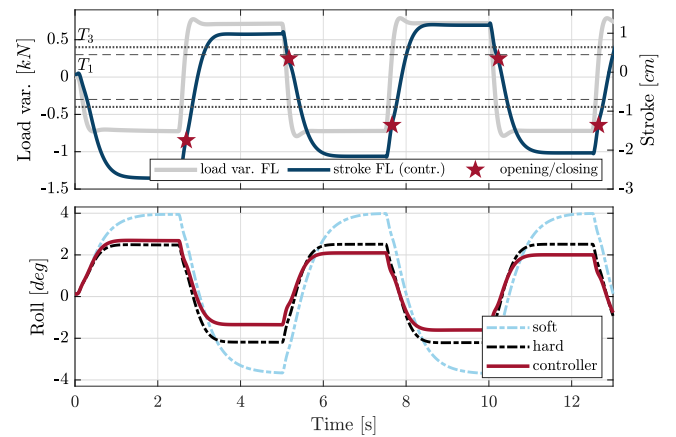}
	\caption{Maneuver inversion management example.}
	\label{fig:inversion_example}
	\end{center}
\end{figure}

Both policies explained in the previous section, \textit{i.e.} kick force avoidance and maneuver inversion management, have been tested on specific experiments, in order to highlight the benefits of the proposed control strategy. The closing threshold is set to 300$N$, corresponding to a critical acceleration equal to 2$m/s^2$ (longitudinal), or equivalently 1$m/s^2$ (lateral).

First, a single maneuver example is considered in Fig. \ref{fig:kick_avoid_example}. Following a medium-intensity brake, the car pitches (bottom plot); as soon as the load transfer force (top plot, left \textit{y}-axis) surpasses the closing threshold the spring stiffens thus reducing the steady-state pitch value with respect to the soft configuration. At the end of the maneuver, the valve opens only when the stroke comes back to the level of the closing instant (top plot, right \textit{y}-axis). In this way, the kick force is avoided and as a consequence the vertical acceleration does not deteriorate.

Second, the case of multiple maneuver inversions is reported in Fig. \ref{fig:inversion_example}. In this example, a chicane-like maneuver, \textit{i.e.} a set of step steers, is considered and the corresponding load variation is reported (top plot, left \textit{y}-axis). It is seen that, following the first maneuver, all maneuvers lead to steady state roll angles which are consistently lower than the full hard configuration, thus outperforming what can be achieved with a passive spring (bottom plot). This is made possible by consecutive opening/closing actions which shift the stroke equilibrium position in a way to minimize the total stroke variation in maneuver inversions (top plot, right \textit{y}-axis). For a clearer understanding, each opening/closing action (\textit{star} marker) schematically corresponds to the sequence of points A and B in Fig. \ref{fig:visualizations} (right).

\subsection{Numerical indexes}
The absolute value of the steady-state angles, \textit{i.e.} computed after the rising transient, evaluate the vehicle handling performance:
\begin{equation}
\begin{aligned}
J_{\phi}=|\phi^{ss}(t)| & & J_{\theta}=|\theta^{ss}(t)| \\
\end{aligned}
\label{handling_indexes}
\end{equation}
In particular, the lower (\ref{handling_indexes}), the better the performance. In order to keep track of the \textit{discomfort} introduced by the opening of the valve, a vertical acceleration index is also considered:
\begin{equation}
J_{z}=\text{max}|A_z(t)|,\text{ }t\in [t^{\text{op}},t^{\text{op}}+\Delta t],
\label{comfort_index}
\end{equation}
where $A_z(t)$ is the COG vertical acceleration and $t^{\text{op}}$ is the opening instant. The lower (\ref{comfort_index}), the better the ride feeling.

\begin{figure}[h]
	\begin{center}	
\includegraphics[width=0.48\textwidth]{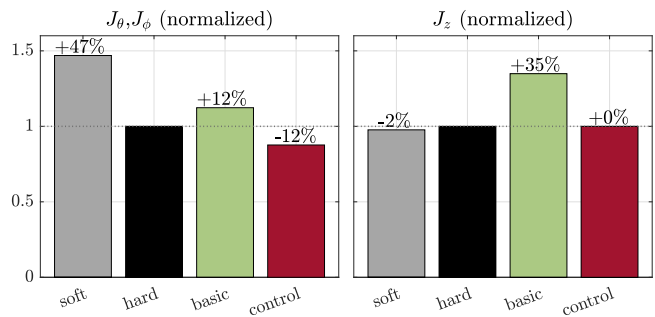}
	\caption{Performance indexes: left - steady state angles in maneuver inversions; right - vertical acceleration peak in standard maneuvers.}
	\label{fig:bar_plot}
	\end{center}
\end{figure}

The performance indexes (normalized with respect to the hard mode) are reported in Fig. \ref{fig:bar_plot}. With reference to the previously introduced tests, the overall improvement in steady-state angles with respect to the full hard configuration is 12\% in the case of a maneuver inversion. Also, the basic stiffening logic is outperformed, since it does not deal with such situations in a smart manner. Also, the proposed logic eliminates the problem of kick force in a standard maneuver (\textit{i.e.} single braking) with no deterioration of the vertical acceleration, conversely to the basic logic which worsens of 35\%. Simulation results hence validate the effectiveness of the proposed control strategy, which makes use of the peculiar features of the multichamber technology in order to improve the performances compared to the basic approach.

\section{Conclusions}

The proposed strategy highlights the benefits introduced by the multichamber architecture in handling-oriented control. The classic \textit{basic} control logic allows to obtain steady state angles comparable the best passive configuration (full hard configuration), despite never outperforming it; however, it simultaneously deteriorates the vertical acceleration due to the kick force effect which characterizes this class of suspensions. On the other hand, the introduction of the kick force mitigation strategy and the maneuver inversion management allows to outperform the classic expectations. Not only the steady state angles in maneuver inversions are lower than what a passive framework is capable of doing, but it also deletes the problem of the kick force in standard maneuvers (\textit{i.e.} maneuvers with no inversions), thus preventing any peaks in vertical acceleration. Future developments concern its implementation on a real vehicle.

\section*{ACKNOWLEDGMENTS}

The authors deeply thank Chiara Martellosio on her advisory contribution for this work.


\begin{table}[thpb]
\begin{center}\caption{Model parameters}
\begin{tabular}{lll}
\hline
\multicolumn{1}{c}{\textbf{Parameter}} & \textbf{Symbol}  & \textbf{Value} \\ \hline
Vehicle sprung mass                    & $M$ $[kg]$       & 2100           \\
Wheelbase                              & $L$ $[m]$        & 3.3            \\
Track                                  & $T$ $[m]$        & 1.6            \\
COG   height                           & $H$ $[m]$        & 0.56           \\
Main chamber nominal   volume          & $V_{main,0}$ $[L]$ & 1.4            \\
Auxiliary chamber volume               & $V_{aux}$ $[L]$    & 1.53           \\
Piston area                            & $A$ $[cm^2]$     & 133            \\
Air polytropic coefficient             & $\gamma$ $[-]$   & 1.4            \\
Damping coefficient                    & $c$ $[Ns/mm]$    & 1.6            \\ \hline
\end{tabular}
\label{table:param}
\end{center}
\end{table}

\bibliographystyle{IEEEtran}
\bibliography{IEEEabrv,mybibfile}

\end{document}